\documentclass[twocolumn]{aastex61}

\newcommand\aastex{AAS\TeX}

\newcommand{\eg}{e.g.}

\newcommand{\figref}[1]{Fig.~\ref{#1}}

\newcommand{\hone}{\ion{H}{1}}

\newcommand{\heone}{\ion{He}{1}}
\newcommand{\hetwo}{\ion{He}{2}}

\newcommand{\cthree}{\ion{C}{3}}
\newcommand{\cfour}{\ion{C}{4}}

\newcommand{\nthree}{\ion{N}{3}}
\newcommand{\nfour}{\ion{N}{4}}
\newcommand{\nfive}{\ion{N}{5}}

\newcommand{\othree}{\ion{O}{3}}
\newcommand{\ofour}{\ion{O}{4}}
\newcommand{\ofive}{\ion{O}{5}}

\newcommand{\sifour}{\ion{Si}{4}}

\newcommand{\sfive}{\ion{S}{5}}

\newcommand{\ebv}{$E($\bv)}

\newcommand{\kms}{km s$^{-1}$}
\newcommand{\logg}{$\log g$}
\newcommand{\msun}{$M_{\sun}$}

\newcommand{\rv}{$R_V$}

\newcommand{\teff}{$T_{\rm eff}$}

\newcommand{\hst}{{\em HST}}

\shorttitle{\aastex\ Y453 in M4}
\shortauthors{Dixon et al.}

\begin{document}

\title{Observations of the Ultraviolet-Bright Star Y453 in the Globular Cluster M4 (NGC~6121)}

\correspondingauthor{William V. Dixon}
\email{dixon@stsci.edu}

\author[0000-0001-9184-4716]{William V. Dixon}
\affil{Space Telescope Science Institute, 3700 San Martin Drive, Baltimore, MD 21218, USA}

\author[0000-0001-7653-0882]{Pierre Chayer}
\affil{Space Telescope Science Institute, 3700 San Martin Drive, Baltimore, MD 21218, USA}

\author[0000-0002-7547-6180]{Marilyn Latour}
\affiliation{Dr. Karl Remeis-Observatory \& ECAP, Astronomical Institute, Friedrich-Alexander University Erlangen-Nuremberg, Sternwartstr. 7, 96049 Bamberg, Germany}

\author[0000-0001-8031-1957]{Marcelo Miguel Miller Bertolami}
\affiliation{Instituto de Astrof\'{i}sica de La Plata, UNLP-CONICET, Paseo del Bosque s/n, 1900 La Plata, Argentina}

\author[0000-0002-8109-2642]{Robert A. Benjamin}
\affiliation{Department of Physics, University of Wisconsin-Whitewater, 800 West Main Street, Whitewater, WI 53190, USA}




\begin{abstract}

We present a spectral analysis of the UV-bright star Y453 in M4.  Model fits to the star's optical spectrum yield \teff\ $\sim 56,000$ K.  Fits to the star's FUV spectrum, obtained with the Cosmic Origins Spectrograph (COS) on board the {\em Hubble Space Telescope}, reveal it to be considerably hotter, with \teff\ $\sim 72,000$ K.  We adopt \teff\ = $72,000 \pm 2000$ K and \logg\ = $5.7 \pm 0.2$ as our best-fit parameters.  Scaling the model spectrum to match the star's optical and near-infrared magnitudes, we derive a mass $M_* = 0.53 \pm 0.24\; M_{\sun}$ and luminosity $\log L/L_{\sun} = 2.84 \pm 0.05$, consistent with the values expected of an evolved star in a globular cluster.  Comparing the star with post-horizontal branch evolutionary tracks, we conclude that it most likely evolved from the blue horizontal branch, departing the AGB before third dredge-up.  It should thus exhibit the abundance pattern (O-poor and Na-rich) characteristic of the second-generation (SG) stars in M4.  We derive the star's photospheric abundances of He, C, N, O, Si, S, Ti, Cr, Fe, and Ni.  CNO abundances are roughly 0.25 dex greater than those of the cluster's SG stars, while the Si and S abundances agree match the cluster values.  Abundances of the iron-peak elements (except for iron itself) are enhanced by 1 to 3 dex.  Rather than revealing the star's origin and evolution, this pattern reflects the combined effects of diffusive and mechanical processes in the stellar atmosphere.

\end{abstract}





\section{Introduction} \label{sec_intro}

The color-magnitude diagrams of Galactic globular clusters frequently include a handful of {luminous, blue} stars.  Called UV-bright stars, these objects are evolving, either from the asymptotic giant branch (AGB) or directly from the extreme horizontal branch (EHB), onto the white-dwarf cooling curve.  Their stellar parameters and photospheric abundances thus provide important constraints on theories of low-mass stellar evolution and white-dwarf formation.

First cataloged by \citet{Cudworth:Rees:1990}, Y453 was identified as a UV-bright star in Ultraviolet Imaging Telescope observations of the globular cluster M4 \citep[NGC 6121;][]{Parise:Stecher:1995}.  The star's optical spectrum was studied by \citet{Moehler:Landsman:Napiwotzki:98}.  Fits to its hydrogen and helium lines yielded an effective temperature \teff\ = 58,800 K, surface gravity \logg\ = 5.15, and helium abundance log N(He)/N(H) = $-0.98$.  These parameters place the star on the 0.546 \msun\ post-early AGB evolutionary track of \citet{Schoenberner:83}, but its derived mass and luminosity ($M_* = 0.16\; M_{\sun}$ and $\log L/L_{\sun} = 2.6$) are inconsistent with this evolutionary scenario.  To better understand this enigmatic object, we have observed Y453 with the Cosmic Origins Spectrograph (COS) on the \emph{Hubble Space Telescope (HST)}.

Our FUV observations are discussed in \S\ \ref{sec_observations}.  Section \ref{sec_optical} presents a reanalysis of the original optical spectrum.  In \S\ \ref{sec_FUV} we return to the FUV spectrum, using it to derive the star's effective temperature and photospheric abundances.  In \S\ \ref{sec_discussion}, we discuss the star's cluster membership, derive its mass and luminosity,  compare its parameters with new evolutionary models, and compare its abundances with the cluster values.  We review our conclusions in \S\ \ref{sec_conclusions}.

\vspace{0.575in}

\section{Observations and Data Reduction}\label{sec_observations}

Y453 was observed with \hst/COS on 2015 February 09 (Program 13721; R. Benjamin, P.I.). A 2048-second exposure, obtained with the G130M grating, extends from 1130 to 1430 \AA\ with a signal-to-noise ratio $S/N \sim 30$ per 7-pixel resolution element.  A 5603-second exposure, obtained with the G160M grating, extends from 1430 to 1770 \AA.  Its $S/N > 50$ at 1430 \AA\ and falls to $\sim 15$ at the longest wavelengths.  The flux- and wavelength-calibrated spectra, processed with CALCOS version 3.0, were retrieved from the Mikulski Archive for Space Telescopes (MAST) on 2015 August 28. {A link to the data is provided here: \dataset[10.17909/T9KC7B]{http://doi.org/10.17909/T9KC7B}.}

\begin{figure}
\plotone{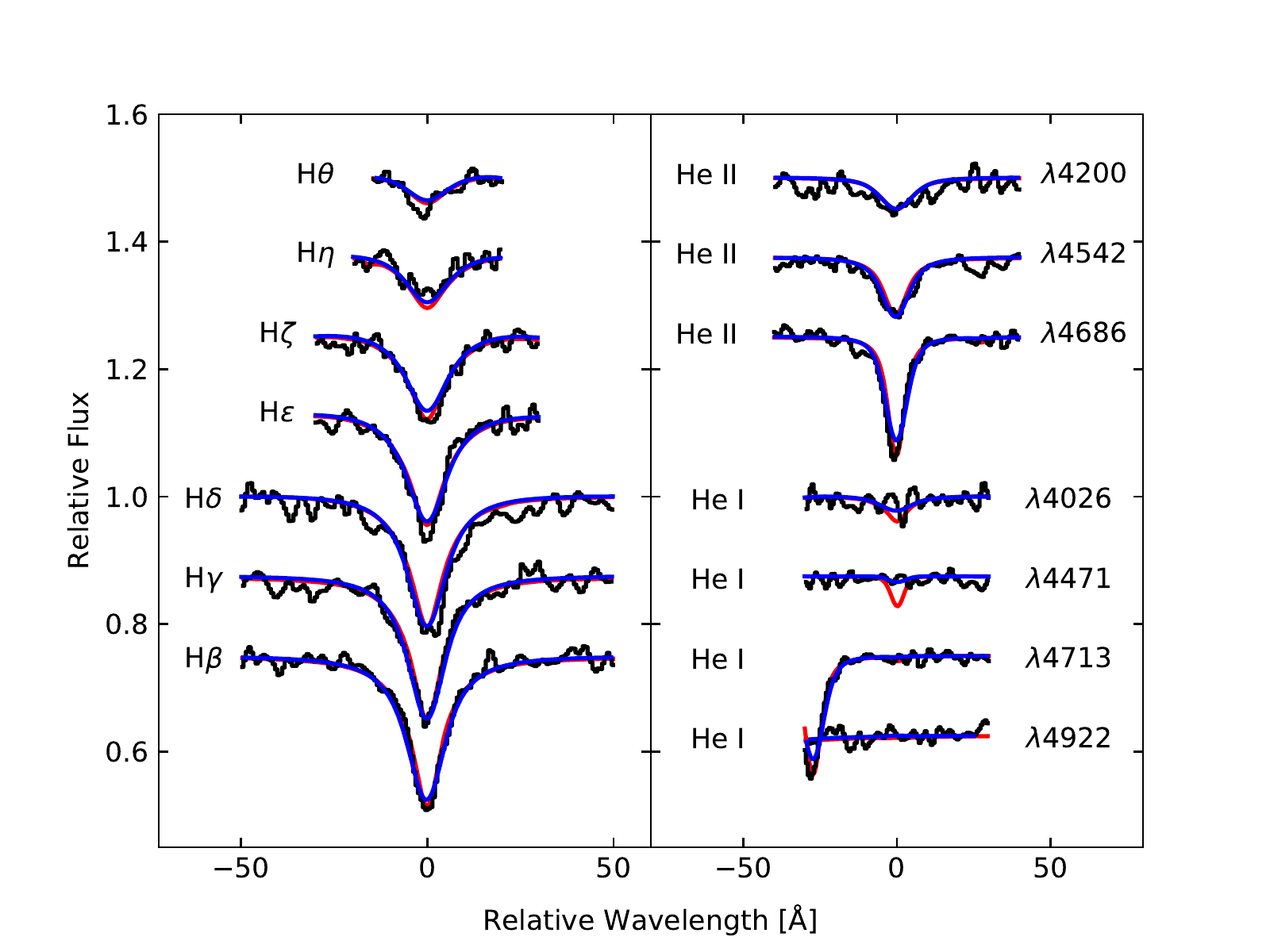}
\caption{{Hydrogen and helium lines in the} optical spectrum of Y453 from \citealt{Moehler:Landsman:Napiwotzki:98} overplotted by our best-fit H+He model (red curve) and best-fit high-metallicity model (ten times solar; blue curve).  The \heone\ $\lambda 4471$ line is not present in the stellar spectrum, but is strong in our H+He models.  The line becomes weaker as the model opacity rises.}
\label{fig_optical}
\end{figure}

\newpage
\section{The Optical Spectrum}\label{sec_optical}

We begin by returning to the optical spectrum of Y453, which was kindly provided by S.~Moehler and is reproduced in \figref{fig_optical}.  Details of the observation, data reduction, and original analysis are presented in \citet{Moehler:Landsman:Napiwotzki:98}.  

We compute a grid of non-local thermodynamic equilibrium (NLTE) stellar atmosphere models using the program TLUSTY \citep{Hubeny:Lanz:95}.  The model atmospheres are composed solely of hydrogen and helium.  The atomic models for \hone, \heone, and \hetwo\ are similar to those that \citet{Lanz:Hubeny:2003} used for computing their grid of O-type stars.  Our grid of models covers  effective temperatures ranging from \teff\ = 52,000 to 70,000~K in steps of 2000~K, gravities from $\log g = 5.0$ to 6.0 in steps of 0.2 dex, and He abundance from log N(He)/N(H) = $-0.6$ to $-2.2$ in steps of 0.4 dex. From that grid of model atmospheres, we compute synthetic spectra using the program SYNSPEC \citep{Hubeny:88}. The synthetic spectra are convolved with a Gaussian and normalized to replicate the observed spectrum.

\citet{Moehler:Landsman:Napiwotzki:98} reported that the resolution of their spectrum is 6.7 \AA, but the cores of the H and He lines are not well fit by models with this resolution.  Experimenting with a range of Gaussian line-spread functions, we find that the line cores are best fit by models with a resolution of 4.2 \AA.  Returning to her observation notes, S. Moehler was able to determine that the detector used in her 1996 observations had 0\farcs34 pixels, so a slit width of 1\arcsec\ should have yielded a spectrum with 5.6 \AA\ resolution.  To achieve a resolution of 4.2 \AA\ would have required atmospheric seeing of 0\farcs8, which is unlikely.  We have accordingly adopted a spectral resolution of 5.6 \AA\ in our analysis.  With this change, our best-fit stellar parameters are \teff\ = $55,218 \pm 405$ K, \logg\ = $5.60 \pm 0.02$, and log N(He)/N(H) = $-1.12 \pm 0.02$.

As we shall see, fits to CNO lines in the star's COS spectrum suggest that \teff\ $\sim$ 72,000~K, considerably higher than the temperature derived from its optical spectrum.  This discrepancy may be a symptom of the Balmer-line problem \citep{Napiwotzki:BalmerProblem:1993, Werner:BalmerProblem:1996}, the inability of models to reproduce simultaneously the full set of Balmer lines with a single set of stellar parameters (\teff\ and \logg).  Apparently, there are sources of opacity in the atmospheres of hot stars that are not included in our models.  An example is the hot subdwarf O star BD +28\degr 4211.  According to \citet{Napiwotzki:BalmerProblem:1993}, fits to the highest Balmer lines (H$\delta$ or H$\epsilon$) yield \teff\ $\sim$ 82,000 K, but fits to the complete Balmer series yield much lower temperatures. \citet{Latour:2015} were able to reproduce the higher temperature by fitting the  star's complete optical spectrum with a stellar model assuming a metallicity ten times solar.

We have extended the \citet{Latour:2015} grid of high-metallicity models to cooler temperatures and fit them to the optical spectrum of Y453.  Assuming a spectral resolution of 5.6 \AA, we derive \teff\ = $55,870 \pm 780$~K, \logg\ = $5.69 \pm 0.04$, and log N(He)/N(H) = $-1.08 \pm 0.04$.  A similar grid of solar-metallicity models yields \teff\ = $54,470 \pm 730$ K, \logg\ = $5.72 \pm 0.05$, and log N(He)/N(H) = $-1.11 \pm 0.04$.  (The metallicity of M4 is [Fe/H] = $-1.16$; \citealt[2010 edition]{Harris:96}.)  In both cases, the spectrum is reasonably well fit.  Most interesting is the \heone\ $\lambda 4471$ line: as shown in \figref{fig_optical}, the feature is not present in the stellar spectrum, but is strong in our H+He models.  It becomes successively fainter as the model metallicity increases, disappearing in the high-metallicity grid.  We conclude that high-metallicity models are better able to reproduce the optical spectrum of Y453, but cannot reproduce the high effective temperature derived from the star's FUV spectrum.  We adopt the stellar parameters derived from the high-metallicity grid as the best fits to the star's optical spectrum.

\begin{figure}
\plotone{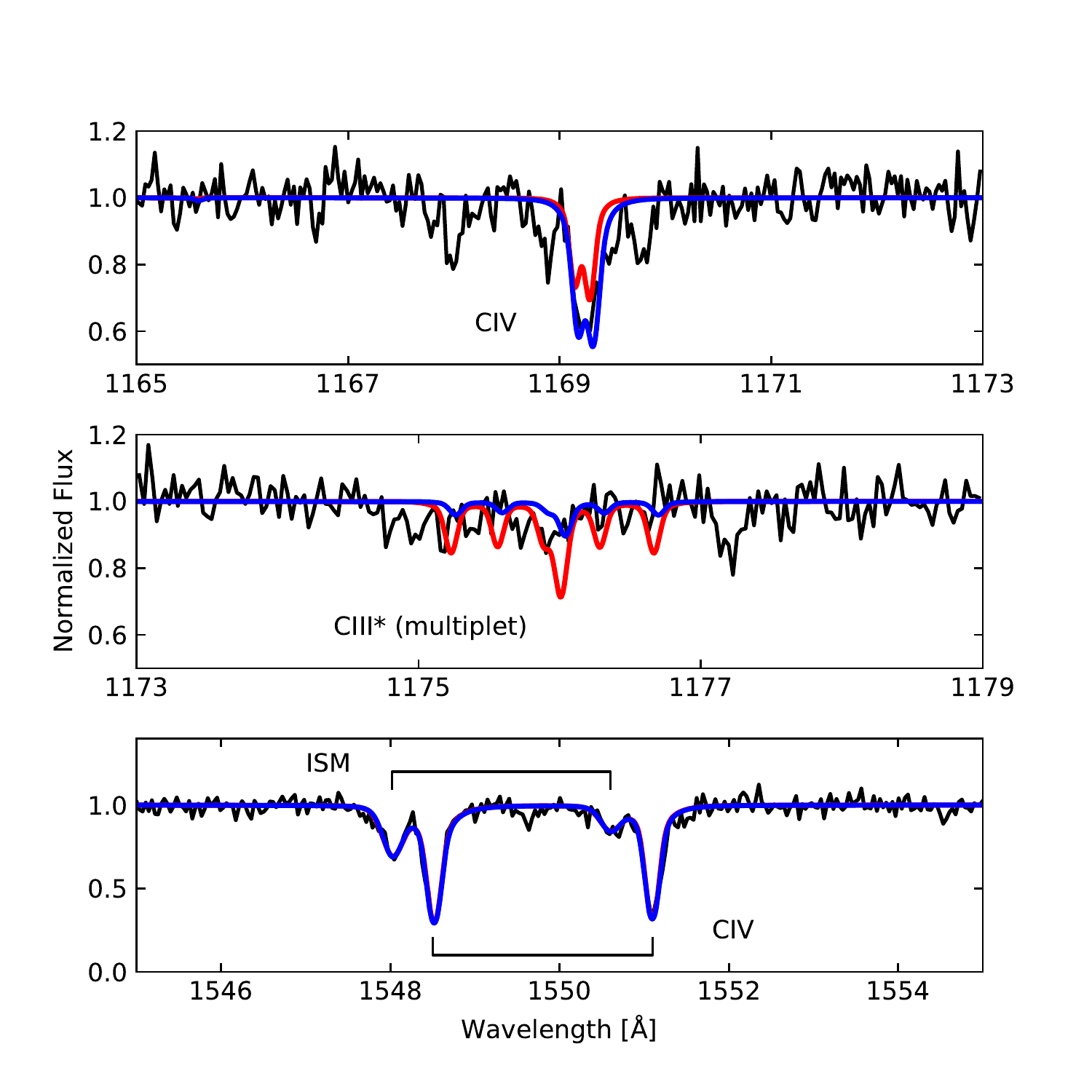}
\caption{Carbon features in the COS spectrum of Y453. The red curve represents a model with \teff\ = 56,000 K; the blue curve is 72,000 K.  Each model has the best-fit carbon abundance for its temperature.  The cooler models over-predict the strength of the \cthree * multiplet and under-predict the \cfour\ lines at 1169 \AA.  The interstellar \cfour\ features are modeled with Gaussian absorption lines.  For this figure, the data are smoothed by 3 pixels.
}
\label{fig_carbon}
\end{figure}

\section{The FUV Spectrum}\label{sec_FUV}

\subsection{Effective Temperature}\label{sec_temperature}

Though its optical spectrum is best fit by models with \teff\ $\sim 56,000$~K, the star's COS spectrum requires models with a higher effective temperature.  We use simultaneous fits of lines from multiple ionization stages to constrain \teff.  We begin by fitting the \cthree\ and \cfour\ features shown in \figref{fig_carbon}.  We construct a grid of models with an atmosphere consisting of H+He+C.  The effective temperature ranges from 50,000 to 77,500~K in steps of 2500 K, and the carbon abundance ranges from log N(C)/N(H) = $-6.8$ to $-3.2$ in steps of 0.4 dex.  In this grid, the surface gravity is fixed at \logg\ = 5.7 and the helium abundance at log N(He)/N(H) = $-1.08$, values derived from the optical spectrum.  Models are resampled to the COS pixel scale and convolved with the COS line-spread function.  The interstellar \cfour\ doublet at 1550 \AA\ is modeled with a pair of Gaussian absorption features whose relative wavelengths and equivalent widths are fixed at the appropriate ratios.  Our best-fit parameters are \teff\ = $68,876 \pm 1451$ K and log N(C)/N(H) = $-4.96 \pm 0.09$.  Quoted errors are purely statistical.  The carbon lines may underestimate the temperature, because there are many weak Co and Ni features near 1175 \AA\ that could masquerade as \cthree * lines.

\begin{figure}
\plotone{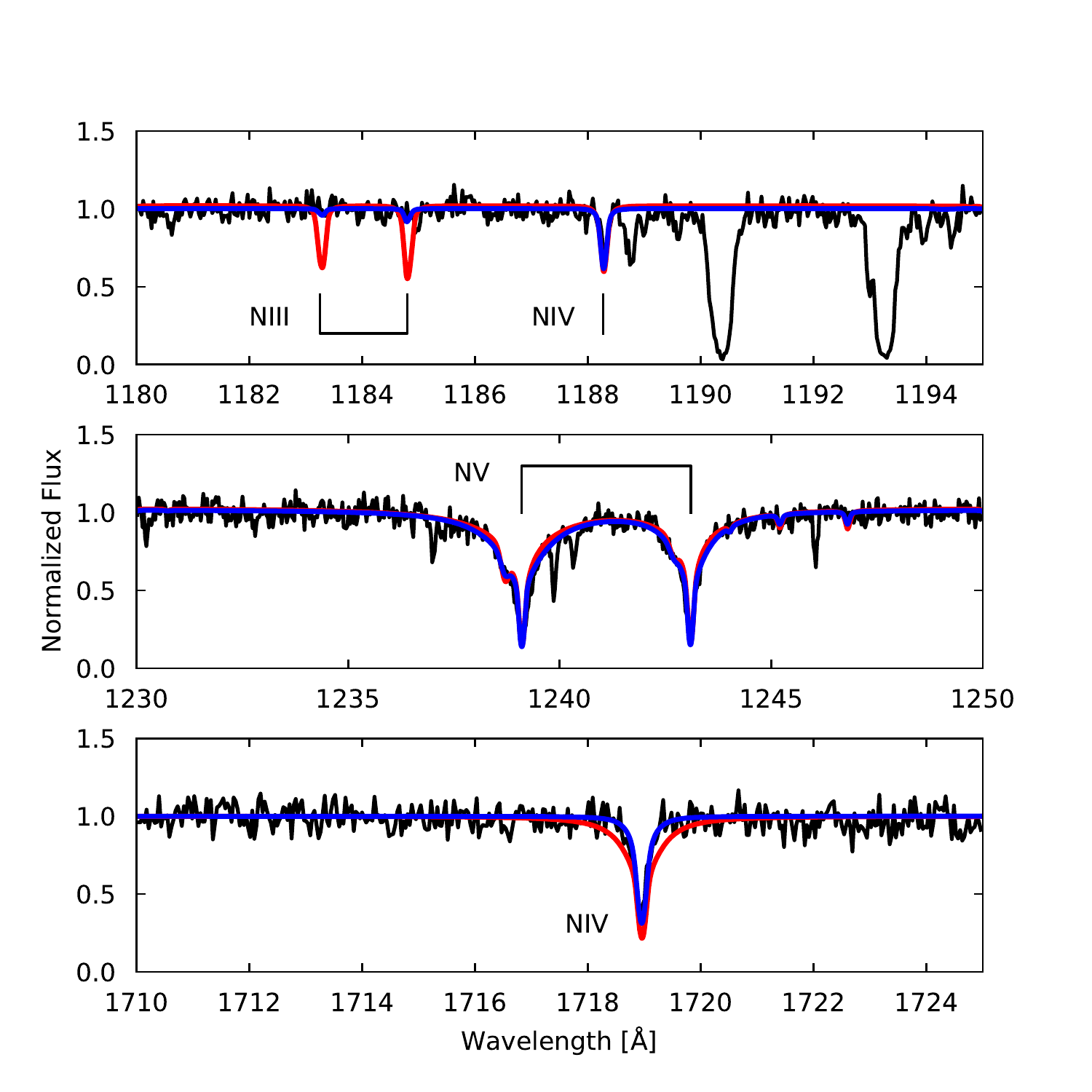}
\caption{Same as \figref{fig_carbon}, but for nitrogen.  The \nfive\ lines are blended with ISM features, which we model using the \cfour\ ISM lines  as templates.  The cooler model over-predicts the strength of the \nthree\ lines.
}
\label{fig_nitrogen}
\end{figure}

\begin{figure}
\plotone{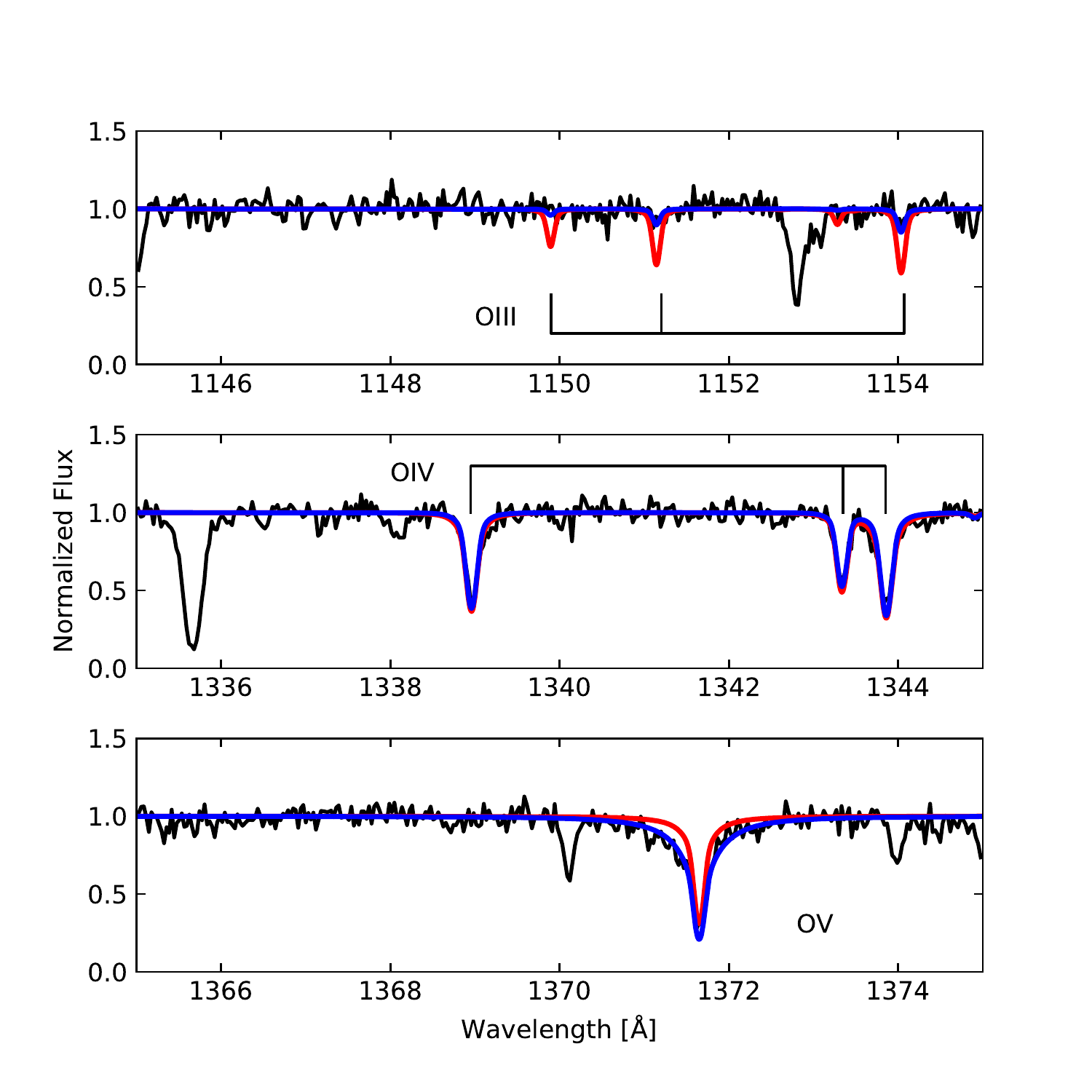}
\caption{Same as \figref{fig_carbon}, but for oxygen.  The cooler model over-predicts the strength of the \othree\ lines.
}
\label{fig_oxygen}
\end{figure}

We repeat this process with the \nthree, \nfour, and \nfive\ lines shown in \figref{fig_nitrogen}.  The star's \nfive\ features are blended with interstellar lines. To account for them, we fit the ISM lines with Gaussians, requiring them to have the same radial velocity (relative to the stellar features) and line width (in units of \kms) as the interstellar \cfour\ lines.  Our best-fit parameters are \teff\ = $71,383 \pm 1064$ K and log N(N)/N(H) = $-3.98 \pm 0.02$.  Repeating this process with the \othree, \ofour, and \ofive\ features shown in \figref{fig_oxygen} yields \teff = $71,884 \pm 808$ K and log N(O)/N(H) = $-3.79 \pm 0.03$.

The effective temperatures derived from the nitrogen and oxygen lines agree within their uncertainties.  The temperature derived from the carbon lines is lower, but it may suffer from contamination of the \cthree * multiplet, as mentioned above.  Thus, we use only the nitrogen and oxygen results.  Their weighted mean is \teff\ = $71,675 \pm 643$ K.  The statistical uncertainties of both the surface gravity and effective temperature are quite small.  To better understand the systematic uncertainties of these parameters and their effects on the photospheric abundances (derived below), we follow \citet{Rauch:2014} in assuming an uncertainty of 0.2 dex in the surface gravity.  Repeating the above analysis assuming \logg\ = 5.5, we derive an effective temperature of \teff\ = $70,087 \pm 698$ K.  We thus adopt as our best-fit parameters \teff\ = $72,000 \pm 2000$ K and \logg\ = $5.7 \pm 0.2$.

\subsection{Surface Gravity and Helium Abundance}\label{sec_gravity}

{We would normally derive the surface gravity and helium abundance from simultaneous fits to the hydrogen and helium lines in the UV spectrum, just as we did in the optical, but the star's Lyman $\alpha$ line is dominated by interstellar absorption and thus unavailable.  The lone helium feature in the COS spectrum is \hetwo\ $\lambda 1640$.  We fit this line with a two-dimensional grid of NLTE H+He models, allowing both the surface gravity and helium abundance to vary freely.  The  best-fit values, \logg\ $\sim$ 6.8 and log N(He)/N(H) $\sim -1.8$ (exact values depend on the details of the fit), are astrophysically implausible (implying a stellar mass of 6.6 \msun) and inconsistent with fits to the optical spectrum.  We suspect that the same opacity effects that alter the depths of the hydrogen and helium lines at optical wavelengths are at work in the FUV, as well.  We will continue to use the optically-derived values of the surface gravity and helium abundance.}

\begin{deluxetable*}{llccc}
\tablecaption{Abundances from Fits to the COS Spectrum \label{tab:abund}}
\tablehead{
\colhead{Species} & \colhead{Lines Fit} & \multicolumn{3}{c}{Abundance} \\
\cline{3-5}
\colhead{} & \colhead{} & \colhead{\teff\ = 72,000 K} & \colhead{\teff\ = 70,000 K} & \colhead{Final} \\
\colhead{} & \colhead{} & \colhead{\logg\ = 5.7} & \colhead{\logg\ = 5.5} & \colhead{Value}
}
\startdata
Carbon          & \cfour\ $\lambda \lambda 1169$,  \cthree * $\lambda \lambda 1175$, \cfour\ $\lambda \lambda 1550$ & $-4.76 \pm 0.04$ & $-4.86 \pm 0.04$ & $-4.76 \pm 0.11$ \\
Nitrogen        &  \nthree\ $\lambda \lambda 1183, 1185$; \nfour\ $\lambda 1188$; & $-4.07 \pm 0.02$ & $-4.08 \pm 0.02$ & $-4.07 \pm 0.02$ \\
 & \nfive\ $\lambda \lambda 1239, 1243$; and \nfour\ $\lambda 1719$ \\
Oxygen          & \othree\ $\lambda \lambda 1149, 1150, 1153$; & $-3.81 \pm 0.03$ & $-3.85 \pm 0.03$ & $-3.81 \pm 0.05$ \\
 & \ofour\ $\lambda \lambda 1338, 1342, 1343$; \ofive\ $\lambda 1371$ \\
Silicon         & \sifour\ $\lambda  \lambda 1394, 1403$ & $-5.02 \pm 0.12$ & $-5.09 \pm 0.13$ & $-5.02 \pm 0.15$ \\
Sulfur          & \sfive\ $\lambda 1502$ & $-5.63 \pm 0.10$ & $-5.71 \pm 0.10$ & $-5.63 \pm 0.13$ \\
Titanium        &\ion{Ti}{5} $\lambda \lambda 1675, 1687$ & $-5.00 \pm 0.36$ & $-5.00 \pm 0.37$ & $-5.00 \pm 0.36$  \\
Chromium        & \ion{Cr}{5} $\lambda \lambda 1490, 1498$ & $-6.99 \pm 0.22$ & $-7.06 \pm 0.20$ & $-6.99 \pm 0.23$  \\
Iron            &  \ion{Fe}{5} and \ion{Fe}{6}, 1373--1390 \AA\ & $-5.54 \pm 0.07$ & $-5.63 \pm 0.08$ & $-5.54 \pm 0.11$ \\
Nickel          &\ion{Ni}{5}, 1310--1330 \AA; \ion{Ni}{6}, 1170--1186 \AA & $-6.01 \pm 0.05$ & $-6.10 \pm 0.04$ & $-6.01 \pm 0.10$\\
\enddata
\tablecomments{Abundances relative to hydrogen: log N(X)/N(H).  For iron and nickel, we fit multiple absorption features in the listed region(s).}
\end{deluxetable*}

\subsection{Photospheric Abundances}\label{sec_abundances}

The COS spectrum of Y453 shows absorption from He, C, N, O, Si, S, Ti, Cr, Fe, and Ni.  Since oxygen and nitrogen are the most abundant metals in the star's photosphere, we determine their abundances first.  To reduce the size of our model grids, we perform the calculation iteratively:  Beginning with a grid of models with H+He+O, we determine the oxygen abundance.  We then generate a grid with H+He+N+O, holding the oxygen abundance fixed, and determine the nitrogen abundance.  Finally, we generate a grid with H+He+N+O, holding the nitrogen abundance fixed, and determine the final oxygen abundance.  For all other species, we generate a grid of models with H+He+N+O+X, where N and O are fixed at their best-fit values.  All of our models assume that log N(He)/N(H) = $-1.08$.  Unless otherwise stated, all species are treated in NLTE.  We perform the entire analysis twice, first assuming \teff\ = 72,000 K and \logg\ = 5.7, then assuming \teff\ = 70,000 K and \logg\ = 5.5.  The difference in the two results is our estimate of the systematic error in our fits.  We add this term and the statistical error in quadrature to compute our final error.  Our results are presented in Table \ref{tab:abund}.  Notes regarding individual species follow.

{\em Carbon:} Omitting the \cfour\ $\lambda \lambda 1550$ doublet from the fit does not significantly change the derived carbon abundance.

{\em Nitrogen:}  While we did not use the line in our fits, we noticed that, in our synthetic spectra, the \nthree\ feature at 1730 \AA\ was in emission, while the data show it in absorption.  We were using the \nthree\ model atom (n3\_25+7lev.dat) that \citet{Lanz:Hubeny:2003} used to compute their grid of O star models.  By adopting a more elaborate model atom (n3\_40+9lev.dat), we were able to reproduce the absorption feature at 1730 \AA. The new model atom includes several energy levels above the lower energy level of the transition (E $> 330,000$ cm$^{-1}$). 

{\em Iron-peak elements:} We lack the model atoms necessary to compute full NLTE models for some iron-peak elements, so use iron, for which we have a complete set of models, to explore NLTE effects in this part of the periodic table.  To this end, we generate a second grid of iron spectra, using an NLTE model with a H+He+N+O composition and the LTE approximation to compute the populations of iron in the stellar atmosphere.  TLUSTY reports that, in an atmosphere with \teff\ = 72,000 K, the dominant ionization state for iron is \ion{Fe}{6}, but the \ion{Fe}{6} lines in our LTE spectra are considerably weaker than those in our NLTE spectra.  Fitting our LTE models to the \ion{Fe}{5} lines in this bandpass yields log N(Fe)/N(H) = $-5.58 \pm 0.07$, a value consistent with that derived from the NLTE models.  Fitting the same models to the \ion{Fe}{6} lines (principally \ion{Fe}{6} $\lambda 1374.6$) yields log N(Fe)/N(H) = $-5.12 \pm 0.13$, an overestimate of roughly 0.4 dex.  \ion{Cr}{5}, \ion{Fe}{5} and \ion{Ni}{5} have ionization energies of 69, 75, and 76 eV, respectively, so we would expect their populations to vary similarly with temperature.  \ion{Ti}{5} is not such a good match: its ionization energy is 99 eV, equal to the value for \ion{Fe}{6}.  If \ion{Ti}{5} behaves like \ion{Fe}{6}, then the titanium abundance derived from \ion{Ti}{5} lines may be overestimated by $\sim 0.4$ dex.

{\em Titanium and Chromium:} For these species, we lack the model atoms necessary to construct a full NLTE model atmosphere.  Instead, we generate an LTE grid of models, just as we did for iron.  We fit only the lines of \ion{Ti}{5} and \ion{Cr}{5} when computing LTE abundances.

{\em Nickel:}  We have model atoms for ionization states \ion{Ni}{1} through \ion{Ni}{6}, but lack a model for \ion{Ni}{7}.  While \ion{Ni}{6} is the dominant ionization state in the line-forming region of the stellar atmosphere, omitting \ion{Ni}{7} from our models will cause us to over-predict the populations of the lower ionization states.  To examine the effect of the missing \ion{Ni}{7} ions, we generate an LTE grid of models, just as we did for iron.  Fitting only the \ion{Ni}{5} features, we derive an abundance log N(Ni)/N(H) = $-5.93 \pm 0.05$, a result consistent with that derived from the NLTE models.

An important systematic uncertainty in our fits is the placement of the stellar continuum.  Our models are incomplete, in that our line lists lack many weak absorption features.  To address this shortcoming, we often set the model continuum slightly higher than the mean level of the observed spectrum.  The idea is that, when the S/N is high, small dips in the spectrum are not noise features, but weak absorption lines not included in the model.  When the S/N is low, these dips are just noise.  As an experiment, we scale the model by a factor of 1.03 relative to the normalized spectrum.  A higher continuum requires deeper absorption features -- and thus higher abundances -- to reproduce the spectrum.  For the iron-peak elements, the best-fit abundances rise by an average of 0.4 dex.  None of the fits reported in Table \ref{tab:abund} employ a rescaled continuum.

\section{Discussion}\label{sec_discussion}

\subsection{Cluster Membership}\label{sec_membership}

Our analysis is predicated on the assumption that Y453 is a member of M4.  Before proceeding, we should confirm that this assumption is correct.  The average heliocentric radial velocity of the cluster is $\langle v \rangle = +71.08 \pm 0.08$ \kms, with a dispersion of 3.97 \kms\ \citep{Malavolta:2015}.  A comparison of the stellar and interstellar features in our COS spectrum yields a heliocentric radial velocity of $+69$ \kms\ for Y453 (B. Wakker 2016, private communication), a value within $1 \sigma$ of the cluster mean. From proper-motion measurements, \citet{Cudworth:Rees:1990} derive a cluster membership probability of 99\% for Y453.  We conclude that Y453 is cluster member.

\subsection{Stellar Mass and Luminosity}\label{sec_mags}

We can derive the star's radius, and from this its mass and luminosity, by comparing its observed and predicted flux.  Any such comparison must take into account dust extinction along the line of sight.  M4 is the globular cluster nearest the sun, yet its extinction is both high and highly variable, owing to its location in the Galactic plane, behind the Sco-Oph cloud complex.  \citet{Hendricks:2012} used a combination of broadband near-infrared (NIR) and optical Johnson-Cousins photometry (specifically, the $B, V, I, J,$ and $K_s$ bands) to study the dust along this line of sight.  They found that the reddening to M4 is well modeled by the extinction curve of \citet*[hereafter CCM]{CCM:89}.  Using this parameterization, they derive a dust-type parameter \rv\ = $3.62 \pm 0.07$, a value considerably higher than the $R_V = 3.1$ commonly assumed for the diffuse ISM.  Across the cluster (roughly 10\arcmin\ $\times$ 10\arcmin), the total range in reddening is about 0.2 magnitudes; the mean extinction is \ebv\ = $0.37 \pm 0.01$.

The spectral irradiance of Y453 has been measured in a range of optical and NIR bands:  $B = 15.914 \pm 0.005$, $V = 15.857 \pm 0.003$ \citep{Mochejska:2002}; $B = 15.852 \pm 0.003$, $V = 15.904 \pm 0.009$, $R_C = 15.792 \pm 0.007$, $J = 15.635 \pm 0.013$ \citep{Libralato:2014}; $J = 15.601 \pm 0.080$, $H = 15.774 \pm 0.149$ from 2MASS \citep{2MASS}; and $G = 15.784 \pm 0.002$ from Gaia Data Release 1 \citep{GaiaDR1}.

\begin{figure}
\plotone{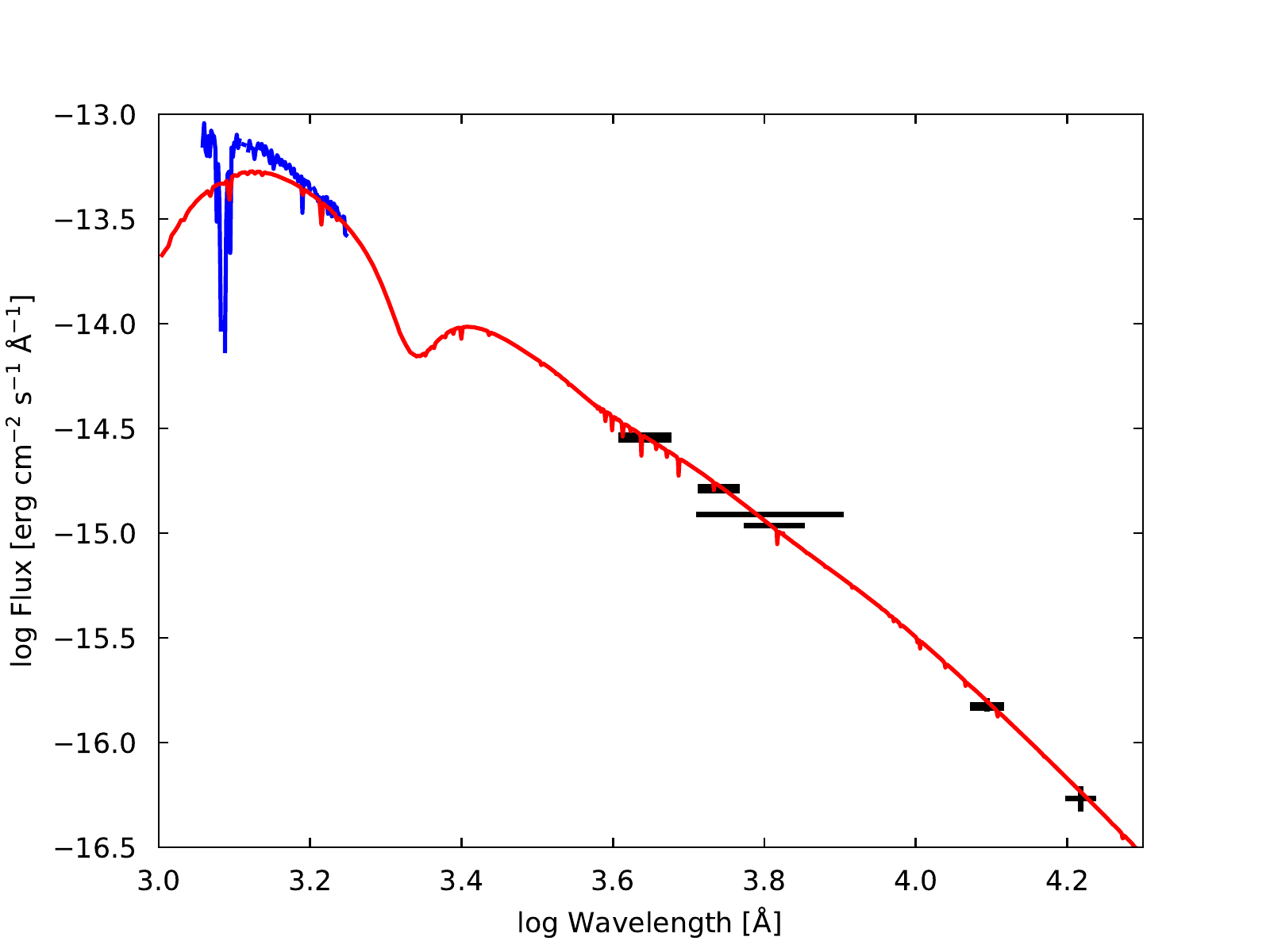}
\caption{Spectral-energy distribution of Y453.  The blue curve is the COS spectrum.  Black crosses are optical and NIR magnitudes from the literature, expressed in units of flux.  The red curve is our best-fit model; scaled to reproduce the optical and NIR data, it under-predicts the flux at FUV wavelengths.  Both the COS and model spectra have been smoothed for this figure.}
\label{fig_fuv_optical}
\end{figure}

To model the stellar continuum, we employ an NLTE model with \teff\ = 72,000 K, \logg\ = 5.7, log N(He)/N(H) = $-1.08$, and the CNO abundances derived in \S\ \ref{sec_abundances}.  Synthetic stellar magnitudes are computed using the Python package Pysynphot \citep{Lim:2015}.  We compute magnitudes relative to the spectrum of Vega, adjusting the zero points of the $B, V,$ and $R_C$ filters as described in Appendix B of \citeauthor{Lim:2015}, but using the {\em V}-band correction of \citet{Bohlin:2007}.  We assume that the zero point for the Gaia {\em G}\/ band is the same as for the {\em V} band, and we take as the zero points for the $J$ and $H$ bands the observed magnitudes of Vega as reported by 2MASS.  We adopt the dust model of \citet{Hendricks:2012} and employ a $\chi^2$ minimization algorithm.  Free parameters in the fit are the scale factor of the model continuum and the extinction parameter \ebv.  
The best-fit model, with a scale factor $\phi = (5.56 \pm 0.05) \times 10^{-23}$ and an extinction \ebv\ = $0.349 \pm 0.002$, is plotted in \figref{fig_fuv_optical}.

In the synthetic spectra generated by SYNSPEC, the flux is expressed in terms of the flux moment, $H_\lambda$.  If the star's radius and distance are known, then the scale factor required to convert the model to the flux at earth is $\phi = 4 \pi (R_* / d)^2$ \citep{Kurucz:79}.  Using the distance $d = 1.82 \pm 0.04$ kpc derived from measurements of three eclipsing binaries in M4 \citep{Kaluzny:2013}, we solve for the stellar radius $R_* = 0.170 \pm 0.004\; R_{\sun}$.  Adopting this value and our best-fit surface gravity (\logg\ = $5.7 \pm 0.2$), we derive a stellar mass $M_* = 0.53 \pm 0.24\; M_{\sun}$.  Finally, combining the stellar radius with our best-fit effective temperature (\teff\ = $72,000 \pm 2000$ K), we derive a stellar luminosity $\log L/L_{\sun} = 2.84 \pm 0.05$.

From a spectroscopic analysis of ten white dwarfs in M4, \citet{Kalirai:2009} found that the mean mass of stars at the tip of the white-dwarf cooling sequence is $0.53 \pm 0.01$ \msun.  If Y453 has evolved past the horizontal branch, as is certainly the case, then it should have a core mass of 0.53 \msun\ and a total mass not much greater.  The derived mass of Y453, $M_* = 0.53 \pm 0.24\; M_{\sun}$, is thus precisely what we would expect of an evolved star in M4.  {The exact agreement of the predicted and derived masses is fortuitous, given the 50\% uncertainty in the derived value, but it is a useful check on the star's surface gravity.  The derived mass scales with the surface gravity (not its logarithm), so is exquisitely sensitive to the adopted value of \logg.}

Recall that our value of the surface gravity, \logg\ = 5.7, was obtained by fitting high-metallicity models to the optical spectrum of Y453.  Had they used similar models, \citet{Moehler:Landsman:Napiwotzki:98} would have derived a stellar mass similar to ours, even with their lower effective temperature.  They wrote, ``In order to obtain a mass of 0.55 \msun, the value of \logg\ would need to be 5.68 instead of 5.15.''

{In \figref{fig_fuv_optical}, the red curve represents the model, and black crosses indicate the optical and NIR magnitudes.  Though the model is scaled to fit only the optical and NIR data, we extend it into the FUV for comparison to the COS spectrum, which is plotted in blue.  The \citet{Hendricks:2012} reddening law for the line of sight to M4 successfully reproduces the observed optical and NIR measurements, but it over-predicts the extinction in the FUV.  Attempts to fit both the COS and optical/NIR data with a single CCM extinction curve have been unsuccessful.  Note that the model spectrum shows no significant Lyman $\alpha$ absorption; the strong line seen in the COS spectrum is interstellar.}

\begin{figure}
\plotone{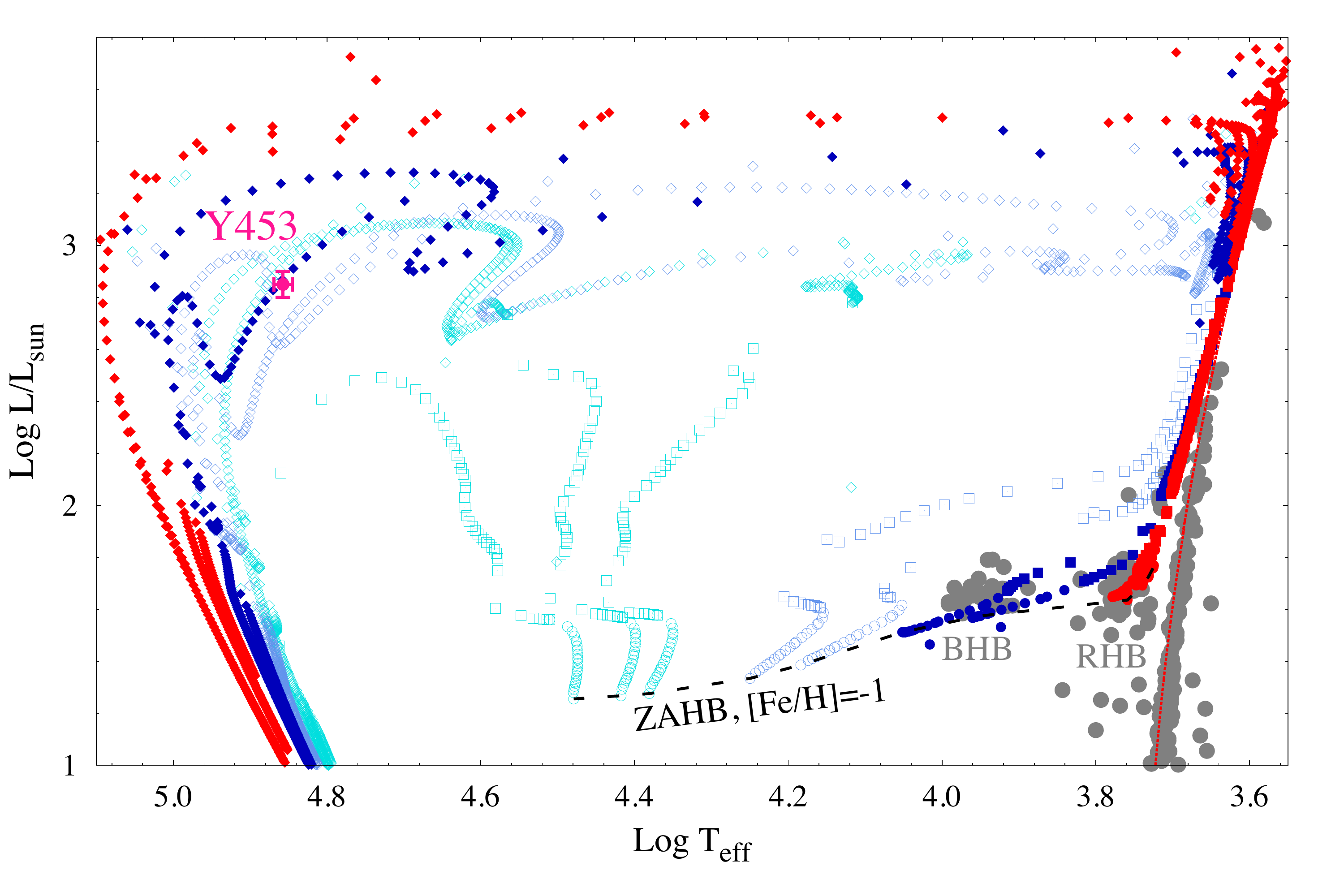}
\caption{Evolutionary tracks for stars similar to those of M4 during and after the horizontal-branch stage. Colors indicate stars that spend their core helium-burning stage on the RHB (red), BHB (blue), and EHB (cyan). Circles are plotted every 10 Myr, squares are plotted every 1 Myr, and diamonds indicate intervals of  20,000 yr.  {Tracks with ZAHB temperatures greater than those of the HB stars in M4 are indicated by open symbols.  Grey points are stars in M4.}  Y453 is indicated by the solid pink circle.}
\label{fig_tracks}
\end{figure}

\subsection{Evolutionary Status}\label{sec_evolution}

To better understand the evolutionary state of Y453, we compare it
with post-HB evolutionary tracks for stars similar to those in M4.
These tracks, shown in \figref{fig_tracks}, represent an extension of the
work presented by \citet{Miller_Bertolami:2016}.  The models are
computed for [Fe/H] = $-1.0$ and a zero-age main-sequence (ZAMS) mass of $M_{\rm ZAMS}=0.85$
\msun, assuming a scaled-solar metal content with
initial abundances $Z_{\rm ZAMS}=0.00172$, $Y_{\rm ZAMS}=0.24844$,
and $X_{\rm ZAMS}=0.74984$.  {The models evolve naturally from the ZAMS, except
on the RGB, where we impose a range of mass-loss rates to ensure population of
the extreme, blue, and red horizontal branches (EHB, BHB, and
RHB, respectively).} Zero-age horizontal-branch (ZAHB) masses are $M_{\rm
ZAHB}= 0.85$, 0.75, 0.70, and 0.65 \msun\ (final masses $M_{\rm WD}=0.555$,
0.545, 0.537, and 0.528 \msun) for the RHB (red points); $M_{\rm ZAHB}=
0.60$, 0.58, 0.55, and 0.53 \msun\ (final masses $M_{\rm WD}=0.518$, 0.513, 0.504, and
0.499 \msun) for the BHB (blue points); and $M_{\rm ZAHB}= 0.50$, 0.495,
and 0.49 \msun\ (final masses $M_{\rm WD}=0.496$, 0.495, and 0.49 \msun)
for the EHB (cyan points). 

{To these evolutionary tracks we have added a sample of M4 stars from the photometry of \citet{Mochejska:2002}.  \citet{Marino:2011} and \citet{Villanova:2012} computed effective temperatures for about two dozen stars that span the HB of M4; we use their results to derive a relation between observed $B-V$ and \teff.  \citet{Malavolta:2014} derived effective temperatures for 2191 stars in M4, but deliberately excluded the HB.  We employ their technique to derive \teff\ for the cluster's giant stars.  Luminosities are computed from {\em V} magnitudes assuming the cluster parameters of \citet{Hendricks:2012} and the bolometric corrections given by \citet{Cox:2000}.  For the purpose of this figure, differential reddening across the cluster is ignored.}

With log \teff\ = 4.86 and $\log L/L_{\sun} = 2.84$, Y453 occupies a region of parameter space
shared by post-EHB stars that evolve directly from the horizontal
branch (cyan points) and post-BHB stars that evolve 
at least partway up the AGB (blue points).  We can reject an
EHB origin for Y453, because M4 does not possess an extreme horizontal
branch (ZAHB stars with $T_{\rm eff} \geq$ 22,000 K).  As shown in \figref{fig_tracks}, the cluster has a
bimodal horizontal branch, well populated on both sides of the RR-Lyrae gap, but its
BHB stars have temperatures $T_{\rm eff} \lesssim 10,000$ K
\citep{Marino:2011, Villanova:2012}.  We conclude
that Y453 is a post-BHB star.  Its low carbon abundance ($N_C/N_O=0.11$) indicates that
the star left the AGB before third dredge-up.

\citet{Villanova:2012} found evidence that the BHB stars in M4 are enriched in helium.
Could their higher helium abundance alter their subsequent evolution?
In fact, the helium enhancement is only about $\Delta$Y = 0.02 relative to the RHB, 
an amount will not significantly affect the BHB stars or their post-BHB evolution. 
In the core helium-burning stage, other uncertainties, among them the size of the convective core, will
dominate.

\begin{deluxetable*}{lccccccc}
\tablecaption{Photospheric Abundances of Y453, M4, and the Sun \label{tab:table}}
\tablehead{
\colhead{Species} & \colhead{Y453} & \multicolumn{2}{c}{M4 RGB}  & & \multicolumn{2}{c}{M4 HB} & \colhead{Sun} \\
\cline{3-4} \cline{6-7}
\colhead{} & \colhead{} & \colhead{FG} & \colhead{SG} & & \colhead{FG/RHB} & \colhead{SG/BHB}
}
\startdata
Helium          & $-1.08 \pm 0.04$ & \nodata & \nodata && \nodata & $-0.99 \pm 0.02$ & $-1.07 \pm 0.01$ \\
Carbon          & $-4.76 \pm 0.10$ & $-4.85 \pm 0.02$ & $-5.01 \pm 0.02$ && \nodata & \nodata & $-3.57 \pm 0.05$ \\
Nitrogen        & $-4.07 \pm 0.02$ & $-5.03 \pm 0.03$ & $-4.39 \pm 0.02$ && \nodata & \nodata & $-4.17 \pm 0.05$ \\
Oxygen          & $-3.81 \pm 0.04$ & $-3.89 \pm 0.03$ & $-4.06 \pm 0.03$ && $-3.79 \pm 0.01$ & $-4.02 \pm 0.02$ & $-3.31 \pm 0.05$ \\
Sodium          & \nodata & $-6.83 \pm 0.02$ & $-6.42 \pm 0.02$ & & $-6.81\pm 0.03$ & $-6.46 \pm 0.02$ & $-5.76 \pm 0.04$ \\
Silicon         & $-5.02 \pm 0.12$ & $-5.10 \pm 0.02$ & $-5.11 \pm 0.02$ && \nodata & \nodata & $-4.49 \pm 0.03$ \\
Sulfur          & $-5.63 \pm 0.13$ & $-5.54 \pm 0.03$ & $-5.54 \pm 0.03$ && \nodata & \nodata & $-4.88 \pm 0.03$ \\
Titanium        & $-5.00 \pm 0.36$ & $-7.85 \pm 0.02$ & $-7.89 \pm 0.01$ && \nodata & \nodata & $-7.05 \pm 0.05$ \\
Chromium        & $-6.99 \pm 0.23$ & $-7.51 \pm 0.02$ & $-7.50 \pm 0.03$ && \nodata & \nodata & $-6.36 \pm 0.04$ \\
Iron            & $-5.54 \pm 0.11$ & $-5.64 \pm 0.01$ & $-5.64 \pm 0.02$ && $-5.64 \pm 0.02$ & $-5.57 \pm 0.02$ & $-4.50 \pm 0.04$ \\
Nickel          & $-6.01 \pm 0.10$ & $-6.88 \pm 0.01$ & $-6.90 \pm 0.01$ && \nodata & \nodata & $-5.78 \pm 0.04$ \\
\enddata
\tablecomments{Abundances relative to hydrogen: log N(X)/N(H). M4 RGB values from \citealt{Villanova:Geisler:2011}; sulfur values from \citealt{Kacharov:2015}.  M4 HB values from \citealt{Marino:2011}; helium value from \citealt{Villanova:2012}.  Solar values from \citealt{Asplund:2009}. FG = first generation; SG = second generation.}
\end{deluxetable*}

\subsection{Photospheric Abundances}\label{sec_levitation}

\begin{figure}
\plotone{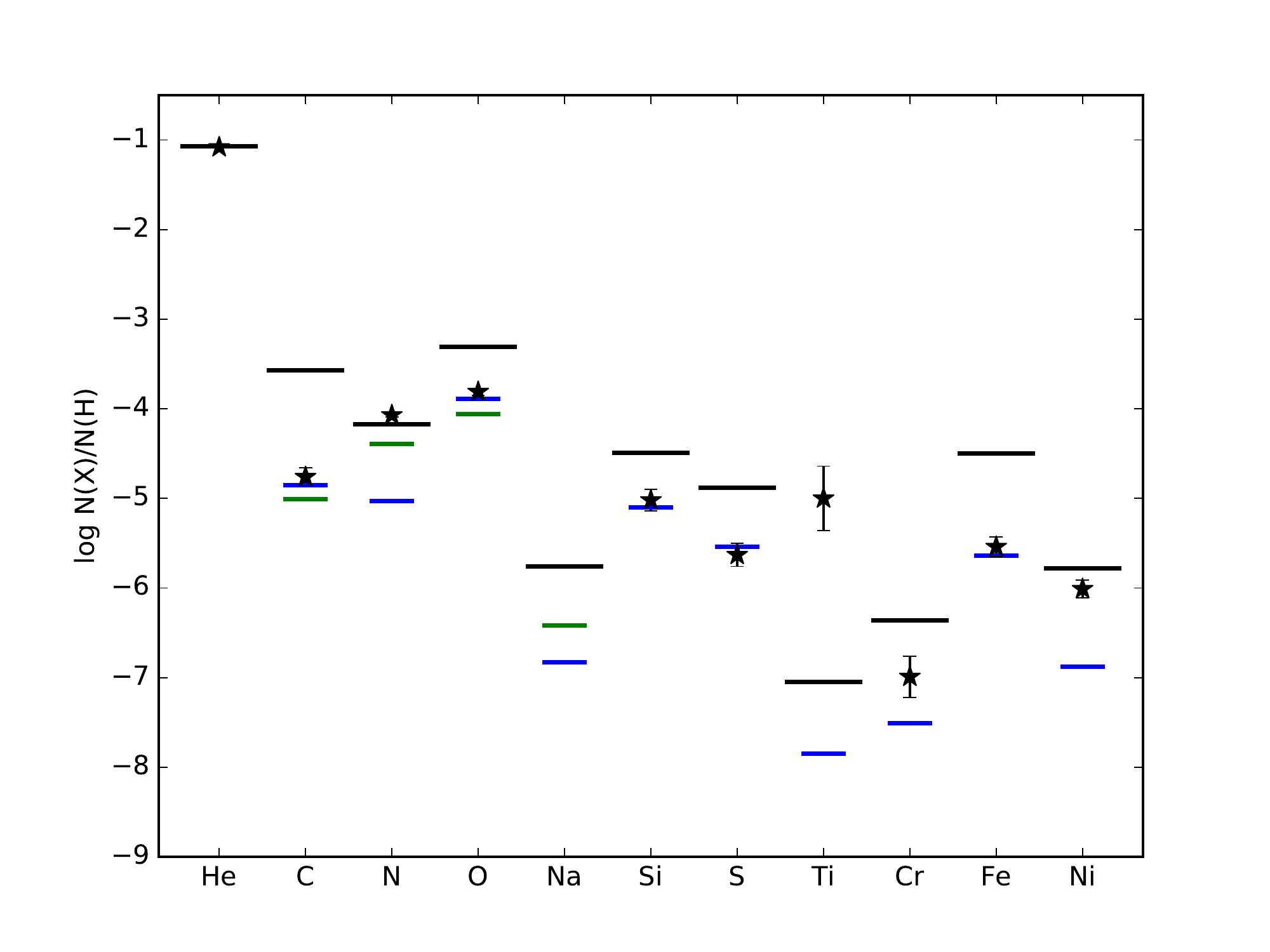}
\caption{Photospheric abundances of Y453 (stars), the RGB stars in M4 (blue and green lines), and the sun (black lines).  Second-generation RGB stars (green lines) are enhanced in N and Na, and depleted in C and O, relative to first-generation RGB stars (blue lines).  The two generations have identical abundances of elements heavier than Na.}
\label{fig_abundance}
\end{figure}

\begin{figure}
\plotone{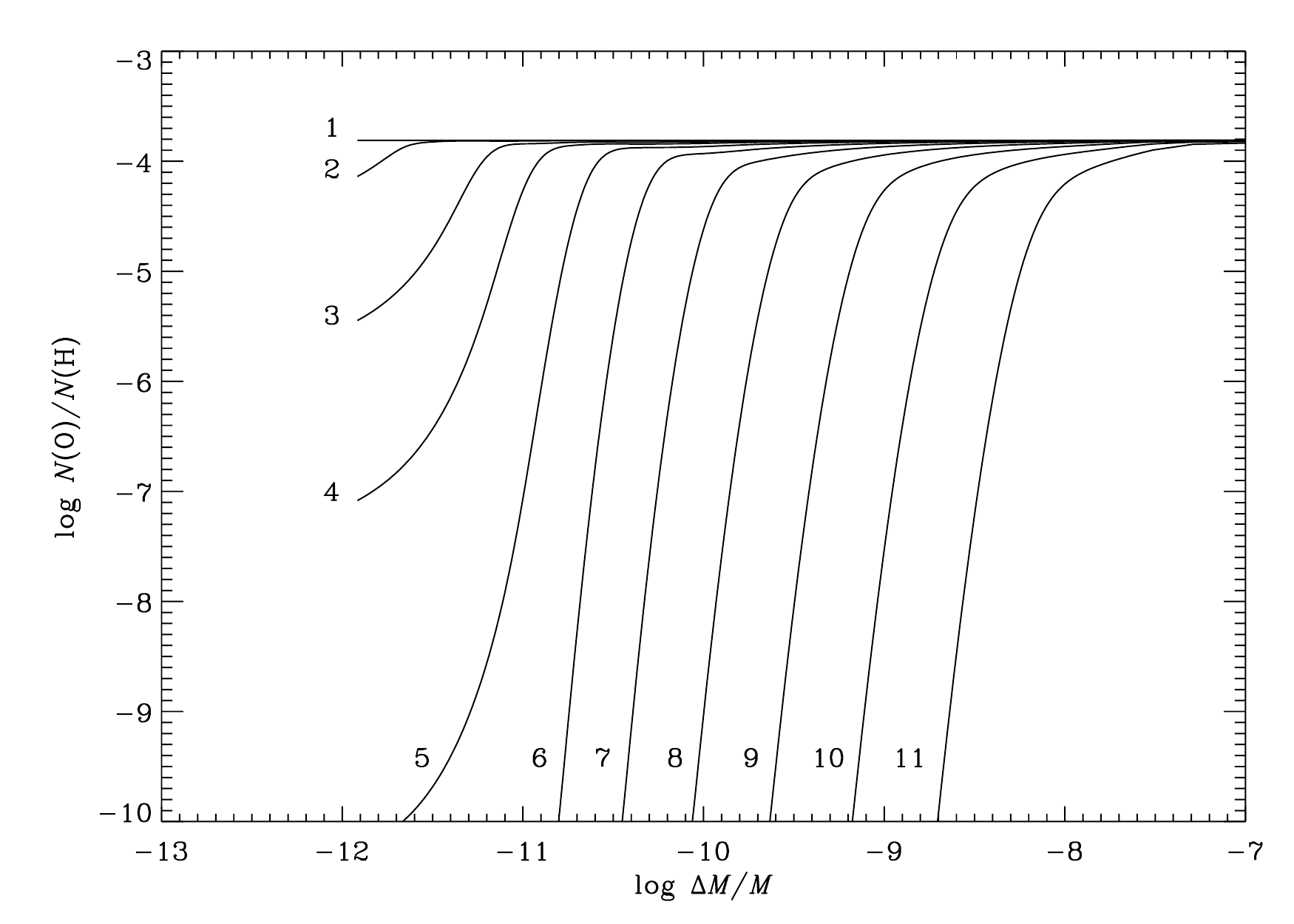}
\caption{Oxygen abundance profiles in a model atmosphere with log \teff\ = 4.86 and \logg\ = 5.6 at eleven time steps. The numbers next to the  abundance profiles indicate the time: 1) 0 yr, 2) 500 yr, 3) 2,500 yr, 4) 5,000 yr, 5) 10,000 yr, 6) 20,000 yr, 7) 40,000 yr, 8) 80,000 yr, 9) 160,000 yr, 10) 320,000 yr, 11) 640,000 yr. The surface of the star is on the left-hand side of the figure. The surface oxygen abundance decreases by about 6 orders of magnitude in less than 10,000 years.}
\label{fig_levitation}
\end{figure}

Galactic globular clusters host multiple stellar populations. First-generation (FG) stars display abundances typical of halo field stars, while second-generation (SG) stars, which may have multiple subpopulations, differ in their abundance of elements affected by proton-capture reactions (\eg, C, N, O, Na).  Models suggest that the second generation formed from gas polluted by material expelled by massive stars of the first generation.  (For details, see the review by \citealt{Gratton:2012}.)  Among RGB stars in M4, second-generation stars are enhanced in N and Na and depleted in C and O relative to first-generation stars \citep{Villanova:Geisler:2011}.  The two populations exhibit the same abundances of $\alpha$ (Mg, Si, S, Ca, Ti) and iron-peak (Cr, Fe, Ni) elements.  The Li and Al abundances and the total C+N+O content are also the same.  Table \ref{tab:table} and \figref{fig_abundance} present the measured abundances of Y453, various subpopulations of M4 \citep{Villanova:Geisler:2011, Kacharov:2015, Marino:2011, Villanova:2012}, and the sun \citep{Asplund:2009}.  

\citet{Marino:2011} found a correlation between the abundances and colors of stars on the HB of M4.  RHB stars exhibit the abundance pattern (O-rich and Na-poor) characteristic of FG stars, while BHB stars exhibit the pattern (O-poor and Na-rich) characteristic of SG stars (Table \ref{tab:table}).  \citet{Villanova:2012} reported that BHB stars are enriched in He, as would be expected of SG stars.  We have concluded that Y453 is a post-BHB star that did not experience third dredge-up, so would expect its abundances to match those of the cluster's SG stars.

In \figref{fig_abundance}, we see that the CNO abundances of Y453 are 0.25 dex greater than those of the second-generation RGB stars, while the Si and S abundances match those of the cluster.  Of the iron-peak elements, only iron itself matches the cluster abundance.  In particular, Ti is over-abundant by roughly three orders of magnitude.  As discussed in \S\ \ref{sec_abundances}, we lack model atoms for Ti, Cr, and the \ion{Ni}{7} ion, but these shortcomings should not result in such large abundance errors.  {We are comparing the abundances of a \teff\ = 72,000 K subdwarf O star derived using NLTE models of FUV lines with those of \teff\ = 5000 K RGB stars derived using LTE models of optical lines (some of them molecular), so systematic differences in the models, the atomic parameters, and the line-formation mechanisms cannot be ignored; however, they are probably not driving the observed discrepancy.  For example, the non-LTE abundance corrections for C, N, O, Na, and Si should be negligible, while those for S and the iron-peak elements are of order 0.1 dex \citep{Asplund:2005}.}

The high luminosity and surface gravity of Y453 suggest that diffusion processes such as gravitational settling are at work in its atmosphere.  To examine this possibility, we have carried out time-dependent diffusion calculations of C, N, O, Si, S, and Fe for two sdO model atmospheres with log \teff\ = 4.84 and \logg\ = 5.6 and 5.8, respectively.  The initial abundances are homogeneous and correspond to the star's observed abundances.  \figref{fig_levitation} shows the evolution of the oxygen abundance as a function of time for the \logg\ = 5.6 model. At the stellar surface, the oxygen abundance decreases by about six orders of magnitude in less than 10,000 years.  For the \logg\ = 5.8 model, the abundance falls even more quickly.  For all species, the abundance drops below log N(X)/N(H) = $-10$ in about 10,000 years.  {From our evolutionary tracks, we estimate that the atmospheric parameters of Y453 have changed little in the past 50,000 years (\figref{fig_tracks}), providing ample time for its photosphere to be stripped of metals.} The fact that we detect any metals, let alone enhancements, suggests that other processes are working to maintain the elements heavier than hydrogen in the atmosphere of the star.

One possibility is that the star's abundances are elevated via radiative levitation.  \citet{Fontaine:2008} were able to reproduce the pulsations observed in the hot sdO star SDSS J160043.6+074802.9 (\teff\ = $71,070 \pm 2725$ K, $\log g = 5.93 \pm 0.11$, and log N(He)/N(H) = $-0.85 \pm 0.08$) by including radiative levitation of iron in their atmospheric model.   (A Fourier analysis of the time-tagged COS data for Y453 reveals no evidence of periodic luminosity variations.)  \citet{Ringat:Rauch:2012} reproduced the general shape of the FUV spectrum of the sdO star EC~11481-2303 (\teff\ = 55,000 K, $\log g = 5.8$, and log N(He)/N(H) = $-2.6$) using a radiative-transport code that models diffusion processes, including radiative levitation. Their best-fit model predicts extreme super-solar abundances for iron and nickel, suggesting that levitation is particularly efficient in that star.  The atmospheric parameters of both stars are similar to those of Y453, so radiative levitation may be at work here, as well. 

Another possibility is that a weak stellar wind strips away the outer layers of the star more quickly than gravitational settling can deplete their metals.  \figref{fig_levitation} shows that, after 50,000 years, gravitational settling affects the outer $10^{-9}$ of the stellar mass.  At lower depths, the abundances are essentially unchanged.  To lose $\Delta M = 5 \times 10^{-10}$ \msun\ via winds in 50,000 years would require a mass-loss rate of $10^{-14}$ \msun\ $\mathrm yr^{-1}$.  There is no evidence of a stellar wind in the spectrum of Y453, but we can set a rough limit on its mass-loss rate.  A COS spectrum of the UV-bright star vZ~1128 in M3 exhibits P~Cygni profiles in a single absorption feature, the \nfive\ $\lambda \lambda 1239, 1243$ doublet \citep{Chayer:2015}.  Model fits to this profile suggest that the wind has a terminal velocity $v_\infty = 380$ \kms\ and mass-loss rate $\dot{M} = 10^{-10}$ \msun\  $\mathrm yr^{-1}$.  For simplicity, we fit the \nfive\ doublet of Y453 using the same wind model used to fit vZ~1128, varying only the mass-loss rate, which we hold constant as function of velocity.  Since our goal is an upper limit on $\dot{M}$, we do not include the ISM features in our model; this choice has little effect, as the wind profile is dominated by material at higher velocities.  We set a $3 \sigma$ upper limit of $\dot{M}q = 1.4 \times 10^{-14}$ \msun\ $\mathrm yr^{-1}$, where $q$ represents the fraction of nitrogen in the form of N$^{+4}$.  TLUSTY predicts that $q \sim 0.8$, so $\dot{M} < 1.75 \times 10^{-14}$ \msun\ $\mathrm yr^{-1}$.  We cannot exclude the possibility that a weak stellar wind is responsible for maintaining the metal abundances seen in the stellar photosphere.

Because gravitational settling would quickly remove all metals from the photosphere, other effects must be at work.  While we cannot quantify their contributions, they are likely to include radiative levitation, a weak stellar wind, and perhaps atmospheric turbulence.

\section{Conclusions}\label{sec_conclusions}

We have performed a spectral analysis of the UV-bright star Y453 in M4.  Fits to the star's optical spectrum with metal-enriched model atmospheres yield \teff\ = $55,870 \pm 780$ K, \logg\ = $5.69 \pm 0.04$.  Fits to the star's COS spectrum reveal it to have an effective temperature \teff\ = $71,675 \pm 643$ K, considerably greater than the optically-derived value.  We adopt \teff\ = $72,000 \pm 2000$ K, \logg\ = $5.7 \pm 0.2$ as our best-fit atmospheric parameters.  We scale the model to match the star's optical and near-infrared magnitudes and derive a stellar mass and luminosity that are consistent with the values expected of an evolved star in M4.  We conclude that the star evolved from the blue horizontal branch, departing the AGB before third dredge-up.  It should thus exhibit the abundance pattern (O-poor and Na-rich) characteristic of SG stars.  Instead, we find that its CNO abundances are roughly 0.25 dex greater than those of the cluster's SG stars, while the Si and S abundances match those of the cluster.  Abundances of the iron-peak elements (except for iron itself) are enhanced by 1 to 3 dex.  {It is likely that the observed abundances of Y453 represent the combined effects of multiple diffusion and mechanical processes within the stellar photosphere.}

While we have resolved the mysteries highlighted by \citet{Moehler:Landsman:Napiwotzki:98} --- the astrophysically implausible values of the star's mass and luminosity --- we have identified one more.  We are unable to resolve the discrepancy between the effective temperatures derived from the star's optical and FUV spectra.  Doing so is likely to require additional optical data with greater resolution and higher signal-to-noise.

\acknowledgments

The authors wish to thank S.\ Moehler for generously providing her optical spectrum of Y453 and for extensive conversations about its analysis.  We thank B.\ Wakker for his analysis of the radial velocity of Y453 and the ISM features in its spectrum.  This work was supported by NASA grant HST-GO-13721.001-A to the University of Wisconsin, Whitewater.  P.C.\ is supported by the Canadian Space Agency under a contract with NRC Herzberg Astronomy and Astrophysics.  {M.L.\ acknowledges support from the Alexander von Humboldt Foundation.  M.M.M.B.\ is partially supported by ANPCyT through grant PICT-2014-2708, by the MinCyT-DAAD bilateral cooperation program through grant DA/16/07, and by a Return Fellowship from the Alexander von Humboldt Foundation.}  This work has made use of NASA's Astrophysics Data System Bibliographic Services (ADS) and the Mikulski Archive for Space Telescopes (MAST), hosted at the Space Telescope Science Institute. STScI is operated by the Association of Universities for Research in Astronomy, Inc., under NASA contract NAS5-26555.  This work has made use of data products from the Two Micron All Sky Survey, which is a joint project of the University of Massachusetts and the Infrared Processing and Analysis Center/California Institute of Technology, funded by the National Aeronautics and Space Administration and the National Science Foundation.  This work has made use of data from the European Space Agency (ESA) mission {\it Gaia}\footnote{\url{http://www.cosmos.esa.int/gaia}}, processed by the {\it Gaia} Data Processing and Analysis Consortium (DPAC)\footnote{\url{http://www.cosmos.esa.int/web/gaia/dpac/consortium}}. Funding for the DPAC has been provided by national institutions, in particular the institutions participating in the {\it Gaia} Multilateral Agreement.

%

\vspace{5mm}
\facilities{HST(COS)}


\software{CALCOS (v3.0), TLUSTY \citep{Hubeny:Lanz:95}, SYNSPEC \citep{Hubeny:88}, Pysynphot \citep{Lim:2015}}


\begin{thebibliography}{}
\expandafter\ifx\csname natexlab\endcsname\relax\def\natexlab#1{#1}\fi
\providecommand{\url}[1]{\href{#1}{#1}}

\bibitem[{{Asplund}(2005)}]{Asplund:2005}
{Asplund}, M. 2005, \araa, 43, 481

\bibitem[{{Asplund} {et~al.}(2009){Asplund}, {Grevesse}, {Sauval}, \&
  {Scott}}]{Asplund:2009}
{Asplund}, M., {Grevesse}, N., {Sauval}, A.~J., \& {Scott}, P. 2009, \araa, 47,
  481

\bibitem[{{Bohlin}(2007)}]{Bohlin:2007}
{Bohlin}, R.~C. 2007, in ASP Conf. Ser. 364, The Future of Photometric,
  Spectrophotometric and Polarimetric Standardization, ed. C.~{Sterken} (San
  Francisco: ASP), 315

\bibitem[{{Brown} \& {Gaia Collaboration}(2016)}]{GaiaDR1}
{Brown}, A.~G.~A., \& {Gaia Collaboration}. 2016, \aap, 595, A2

\bibitem[{{Cardelli} {et~al.}(1989){Cardelli}, {Clayton}, \& {Mathis}}]{CCM:89}
{Cardelli}, J.~A., {Clayton}, G.~C., \& {Mathis}, J.~S. 1989, \apj, 345, 245

\bibitem[{{Chayer} {et~al.}(2015){Chayer}, {Dixon}, {Fullerton},
  {Ooghe-Tabanou}, \& {Reid}}]{Chayer:2015}
{Chayer}, P., {Dixon}, W.~V., {Fullerton}, A.~W., {Ooghe-Tabanou}, B., \&
  {Reid}, I.~N. 2015, \mnras, 452, 2292

\bibitem[{{Cox}(2000)}]{Cox:2000}
{Cox}, A.~N. 2000, {Allen's Astrophysical Quantities, 4th ed.} (New York: AIP
  Press; Springer), 388

\bibitem[{{Cudworth} \& {Rees}(1990)}]{Cudworth:Rees:1990}
{Cudworth}, K.~M., \& {Rees}, R. 1990, \aj, 99, 1491

\bibitem[{{Fontaine} {et~al.}(2008){Fontaine}, {Brassard}, {Green}, {Chayer},
  {Charpinet}, {Andersen}, \& {Portouw}}]{Fontaine:2008}
{Fontaine}, G., {Brassard}, P., {Green}, E.~M., {et~al.} 2008, \aap, 486, L39

\bibitem[{{Gratton} {et~al.}(2012){Gratton}, {Carretta}, \&
  {Bragaglia}}]{Gratton:2012}
{Gratton}, R.~G., {Carretta}, E., \& {Bragaglia}, A. 2012, \aapr, 20, 50

\bibitem[{{Harris}(1996)}]{Harris:96}
{Harris}, W.~E. 1996, \aj, 112, 1487

\bibitem[{{Hendricks} {et~al.}(2012){Hendricks}, {Stetson}, {VandenBerg}, \&
  {Dall'Ora}}]{Hendricks:2012}
{Hendricks}, B., {Stetson}, P.~B., {VandenBerg}, D.~A., \& {Dall'Ora}, M. 2012,
  \aj, 144, 25

\bibitem[{{Hubeny}(1988)}]{Hubeny:88}
{Hubeny}, I. 1988, Comput. Phys. Comm., 52, 103

\bibitem[{{Hubeny} \& {Lanz}(1995)}]{Hubeny:Lanz:95}
{Hubeny}, I., \& {Lanz}, T. 1995, \apj, 439, 875

\bibitem[{{Kacharov} {et~al.}(2015){Kacharov}, {Koch}, {Caffau}, \&
  {Sbordone}}]{Kacharov:2015}
{Kacharov}, N., {Koch}, A., {Caffau}, E., \& {Sbordone}, L. 2015, \aap, 577,
  A18

\bibitem[{{Kalirai} {et~al.}(2009){Kalirai}, {Saul Davis}, {Richer},
  {Bergeron}, {Catelan}, {Hansen}, \& {Rich}}]{Kalirai:2009}
{Kalirai}, J.~S., {Saul Davis}, D., {Richer}, H.~B., {et~al.} 2009, \apj, 705,
  408

\bibitem[{Kaluzny {et~al.}(2013)Kaluzny, Thompson, Rozyczka, Dotter,
  Krzeminski, Pych, Rucinski, Burley, \& Shectman}]{Kaluzny:2013}
Kaluzny, J., Thompson, I.~B., Rozyczka, M., {et~al.} 2013, \aj, 145, 43

\bibitem[{{Kurucz}(1979)}]{Kurucz:79}
{Kurucz}, R.~L. 1979, \apjs, 40, 1

\bibitem[{{Lanz} \& {Hubeny}(2003)}]{Lanz:Hubeny:2003}
{Lanz}, T., \& {Hubeny}, I. 2003, \apjs, 146, 417

\bibitem[{{Latour} {et~al.}(2015){Latour}, {Fontaine}, {Green}, \&
  {Brassard}}]{Latour:2015}
{Latour}, M., {Fontaine}, G., {Green}, E.~M., \& {Brassard}, P. 2015, \aap,
  579, A39

\bibitem[{{Libralato} {et~al.}(2014){Libralato}, {Bellini}, {Bedin}, {Piotto},
  {Platais}, {Kissler-Patig}, \& {Milone}}]{Libralato:2014}
{Libralato}, M., {Bellini}, A., {Bedin}, L.~R., {et~al.} 2014, \aap, 563, A80

\bibitem[{{Lim} {et~al.}(2015){Lim}, {Diaz}, \& {Laidler}}]{Lim:2015}
{Lim}, P.~L., {Diaz}, R.~I., \& {Laidler}, V. 2015, {PySynphot User’s Guide}
  (Baltimore: STScI)

\bibitem[{{Malavolta} {et~al.}(2015){Malavolta}, {Piotto}, {Bedin}, {Sneden},
  {Nascimbeni}, \& {Sommariva}}]{Malavolta:2015}
{Malavolta}, L., {Piotto}, G., {Bedin}, L.~R., {et~al.} 2015, \mnras, 454, 2621

\bibitem[{{Malavolta} {et~al.}(2014){Malavolta}, {Sneden}, {Piotto}, {Milone},
  {Bedin}, \& {Nascimbeni}}]{Malavolta:2014}
{Malavolta}, L., {Sneden}, C., {Piotto}, G., {et~al.} 2014, \aj, 147, 25

\bibitem[{{Marino} {et~al.}(2011){Marino}, {Villanova}, {Milone}, {Piotto},
  {Lind}, {Geisler}, \& {Stetson}}]{Marino:2011}
{Marino}, A.~F., {Villanova}, S., {Milone}, A.~P., {et~al.} 2011, \apj, 730,
  L16

\bibitem[{{Miller Bertolami}(2016)}]{Miller_Bertolami:2016}
{Miller Bertolami}, M.~M. 2016, \aap, 588, A25

\bibitem[{{Mochejska} {et~al.}(2002){Mochejska}, {Kaluzny}, {Thompson}, \&
  {Pych}}]{Mochejska:2002}
{Mochejska}, B.~J., {Kaluzny}, J., {Thompson}, I., \& {Pych}, W. 2002, \aj,
  124, 1486

\bibitem[{{Moehler} {et~al.}(1998){Moehler}, {Landsman}, \&
  {Napiwotzki}}]{Moehler:Landsman:Napiwotzki:98}
{Moehler}, S., {Landsman}, W., \& {Napiwotzki}, R. 1998, \aap, 335, 510

\bibitem[{{Napiwotzki}(1993)}]{Napiwotzki:BalmerProblem:1993}
{Napiwotzki}, R. 1993, \actaa, 43, 343

\bibitem[{{Parise} {et~al.}(1995){Parise}, {Stecher}, \& {UIT
  Team}}]{Parise:Stecher:1995}
{Parise}, R., {Stecher}, T.~P., \& {UIT Team} 1995, \baas, 27, 836

\bibitem[{{Rauch} {et~al.}(2014){Rauch}, {Rudkowski}, {Kampka}, {Werner},
  {Kruk}, \& {Moehler}}]{Rauch:2014}
{Rauch}, T., {Rudkowski}, A., {Kampka}, D., {et~al.} 2014, \aap, 566, A3

\bibitem[{{Ringat} \& {Rauch}(2012)}]{Ringat:Rauch:2012}
{Ringat}, E., \& {Rauch}, T. 2012, in ASP Conf. Ser. 452, Fifth Meeting on Hot
  Subdwarf Stars and Related Objects, ed. D.~{Kilkenny}, C.~S. {Jeffery}, \&
  C.~{Koen} (San Francisco: ASP), 71

\bibitem[{{Sch\"{o}nberner}(1983)}]{Schoenberner:83}
{Sch\"{o}nberner}, D. 1983, \apj, 272, 708

\bibitem[{{Skrutskie} {et~al.}(2006)}]{2MASS}
{Skrutskie}, M.~F., {et~al.} 2006, \aj, 131, 1163

\bibitem[{{Villanova} \& {Geisler}(2011)}]{Villanova:Geisler:2011}
{Villanova}, S., \& {Geisler}, D. 2011, \aap, 535, A31

\bibitem[{{Villanova} {et~al.}(2012){Villanova}, {Geisler}, {Piotto}, \&
  {Gratton}}]{Villanova:2012}
{Villanova}, S., {Geisler}, D., {Piotto}, G., \& {Gratton}, R.~G. 2012, \apj,
  748, 62

\bibitem[{{Werner}(1996)}]{Werner:BalmerProblem:1996}
{Werner}, K. 1996, \apjl, 457, L39

\end{thebibliography}
\end{document}